\documentclass[
letter,
twocolumn,
superscriptaddress,
amsmath,
amssymb,
prl,
nofootinbib
showkeys,
10pt,
floatfix,
nobibnotes,
aps,
fltpage
]{revtex4-2}

\usepackage{latexsym}
\usepackage{amsmath}
\usepackage{amssymb}
\usepackage[T1]{fontenc}
\usepackage[open]{bookmark}
\usepackage{hyperref}
\hypersetup{colorlinks=true,allcolors=blue}
\usepackage{upgreek}

\usepackage{orcidlink}

\usepackage{dcolumn}
\usepackage{bm}
\usepackage{xcolor}
\usepackage{hyperref}
\usepackage{cleveref}
\usepackage{float}
\usepackage{svg}
\usepackage{amssymb}
\usepackage{scalerel}
\usepackage{caption}
\usepackage{wasysym}
\usepackage{lipsum}
\usepackage{enumitem}

\providecommand{\pnl}[1]{{\textcolor{black}{(#1)}}}

\newcommand{\red}[1]{\textcolor{black}{#1}}

\makeatletter
\pretocmd\frontmatter@keys@format{\addvspace{20\p@}}{}{}
\makeatother

\usepackage{svg}
\usepackage{cleveref}
\usepackage{siunitx}
\usepackage{caption}
\usepackage{multirow}
\usepackage{makecell}
\usepackage{tabularray}
\usepackage{makecell} 
\usepackage{array} 
\usepackage{graphicx} 
\usepackage{booktabs}

\usepackage[normalem]{ulem}
\usepackage{soul}

\begin{document}

\title{Spectral Anisotropy in Transition Radiation from Biaxial Media}

\author{Sven Ebel\,\orcidlink{0009-0005-3224-6413}}
\affiliation{POLIMA---Center for Polariton-driven Light--Matter Interactions, University of Southern Denmark, Campusvej 55, DK-5230 Odense M, Denmark}
\affiliation{Institute of Photonics and Quantum Sciences, SUPA, Heriot--Watt University, Edinburgh EH14 4AS, United Kingdom}

\author{Bibi Mary Francis\,\orcidlink{0000-0003-1282-9564}}
\affiliation{Institute of Photonics and Quantum Sciences, SUPA, Heriot--Watt University, Edinburgh EH14 4AS, United Kingdom}

\author{Min~Seok~Jang\,\orcidlink{0000-0002-5683-1925}}
\affiliation{School of Electrical Engineering, Korea Advanced Institute of Science and Technology (KAIST), Daejeon 34141, Korea}

\author{Brian D. Gerardot\,\orcidlink{0000-0002-0279-898X}}
\affiliation{Institute of Photonics and Quantum Sciences, SUPA, Heriot--Watt University, Edinburgh EH14 4AS, United Kingdom}

\author{N.~Asger~Mortensen\,\orcidlink{0000-0001-7936-6264}}
\affiliation{POLIMA---Center for Polariton-driven Light--Matter Interactions, University of Southern Denmark, Campusvej 55, DK-5230 Odense M, Denmark}
\affiliation{D-IAS---Danish Institute for Advanced Study, University of Southern Denmark, Campusvej 55, DK-5230 Odense M, Denmark}

\author{Mauro Brotons-Gisbert\,\orcidlink{0000-0001-7254-8292}}
\affiliation{Institute of Photonics and Quantum Sciences, SUPA, Heriot--Watt University, Edinburgh EH14 4AS, United Kingdom}

\author{Sergii Morozov\,\orcidlink{0000-0002-5415-326X}}
\email{Corresponding author: semo@mci.sdu.dk}
\affiliation{POLIMA---Center for Polariton-driven Light--Matter Interactions, University of Southern Denmark, Campusvej 55, DK-5230 Odense M, Denmark}

\date{\today}

\begin{abstract}
\vspace{0.0cm}
\textbf{Abstract.} 
In anisotropic optical media, the electromagnetic response depends on the orientation of the optical field relative to the material's principal dielectric axes. 
While this direction dependence is well understood in conventional optics, it should also influence light-generation processes driven by free electrons.
Here, we experimentally observe spectrally anisotropic transition radiation from biaxial van~der~Waals crystals. 
Using cathodoluminescence spectroscopy on germanium sulphide (GeS) and molybdenum oxydichloride (MoOCl$_2$) crystals, we show that the transition-radiation spectra differ along the principal in-plane optical axes.
To describe this spectral anisotropy, we develop a thin-film transition-radiation model that reproduces the experimental observations. 
Our results demonstrate that transition radiation is a sensitive probe of the axis-dependent dielectric response of biaxial optical media and suggest that optical anisotropy can provide an additional degree of freedom for free electron-driven spectroscopy, radiation sources, and transition-radiation-based diagnostics.

\vspace{0.3cm}
\end{abstract}

\maketitle

\section{Introduction}

In anisotropic optical media, the dielectric response depends on the orientation of the optical field relative to the material axes and is therefore described by a dielectric tensor. 
This gives rise to direction- and polarization-dependent optical responses, including birefringence and anisotropic light propagation~\cite{LandauLifshitz1984}. 
Biaxial media, with three inequivalent principal dielectric axes, represent the most general case of linear optical anisotropy~\cite{BornWolf:1999:Book}. 
Such strongly direction-dependent responses arise naturally in low-symmetry layered crystals, making van~der~Waals (vdW) materials particularly attractive platforms for studying biaxial optical phenomena in thin, well-defined geometries~\cite{Ma2018}.
Their anisotropic dielectric response should likewise influence light-generation processes driven by free electrons.

Cathodoluminescence (CL) spectroscopy combines electron-beam excitation with optical detection, providing access to the spectral, spatial, angular, and polarization properties of emitted light~\cite{Polman2019,RoquesCarmes2023}. 
Owing to its nanoscale localization and broadband excitation, CL is a versatile probe of optical responses in photonic, plasmonic, and low-dimensional materials, including anisotropic systems~\cite{Rich1997,Akbari2022,Abdi2025}.
The CL signal can originate from both incoherent electron-hole recombination and coherent radiation driven directly by the electromagnetic field of the moving electron~\cite{Brenny2014}. 
One of the most general coherent emission mechanisms is transition radiation, which is generated when a charged particle crosses an interface between media with different dielectric responses~\cite{Ginzburg1996}. 
Unlike Cherenkov or Smith--Purcell radiation, transition radiation does not require a threshold electron velocity or a periodic structure and can therefore occur broadly at material interfaces~\cite{Chen2023}.
Its spectral and angular properties are determined by the electron velocity, interface geometry, and dielectric response of the material.

Transition radiation has been widely studied in isotropic media and, more recently, in structured and dispersive photonic environments where its spectrum and directionality can be strongly modified~\cite{Yu2019,Chen_PRL_2023,Chen_Brewster_2023,Wang2025}. 
In thin films, radiation generated at the entrance and exit interfaces can interfere, producing resonances whose spectral positions depend on the film thickness, dielectric function, and electron velocity~\cite{Yamamoto1996,Lin2018,Ebel_TREX_2026}. How this response is modified when the dielectric function becomes tensorial is much less clear, particularly for biaxial media with inequivalent in-plane optical axes. 
Understanding this regime is important both for interpreting CL from anisotropic materials and for exploiting crystal orientation as an additional degree of freedom in transition-radiation spectroscopy.

Here, we experimentally investigate spectrally anisotropic transition radiation from suspended thin van~der~Waals crystals. 
We introduce polarization-selective CL measurements in parabolic mirror geometry for characterization of in-plane optical anisotropy.
We demonstrate the orientation-independent CL response expected for in-plane isotropic vdW crystals using tungsten disulphide (WS$_2$) and gallium sulphide (GaS) as reference systems. 
We then investigate the biaxial crystals germanium sulphide (GeS) and molybdenum oxydichloride (MoOCl$_2$), where rotation of the crystal relative to the detected polarization produces pronounced changes in the transition-radiation spectrum. 
We develop a theoretical framework that relates the anisotropic transition-radiation spectrum to the principal in-plane dielectric responses of the material.
By further varying the electron energy, we show that the spectral position and tunability of the transition-radiation resonances depend on the principal-axis dielectric response. 
Our results show that transition radiation is sensitive to the orientation-dependent dielectric response of biaxial media, enabling nanoscale spectroscopy of anisotropic optical properties with potential applications in free-electron light sources, particle detection, and electron-beam diagnostics.

\section{Results}

\begin{figure}[t!]
\includegraphics[width=1\linewidth]{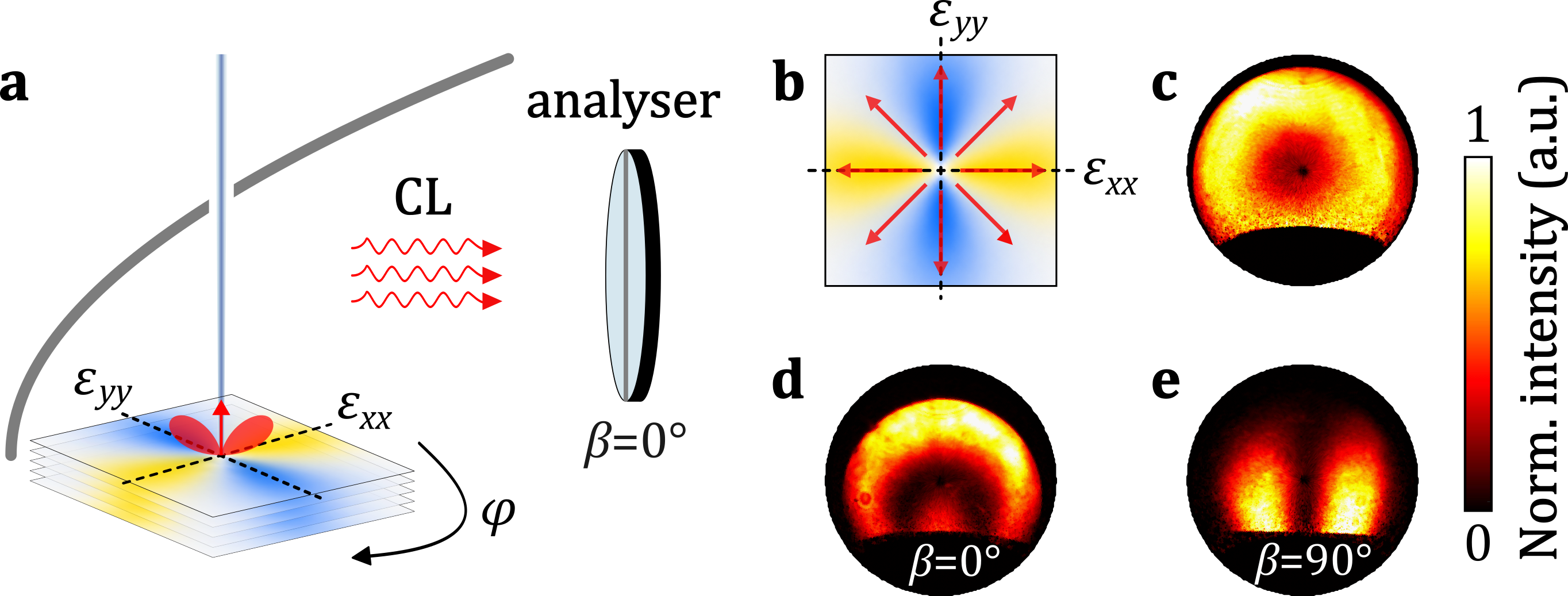}
\caption{\textbf{Polarization-selective detection of anisotropic transition radiation.}
\pnl{a} Schematic of the polarization-selective CL geometry, where transition radiation from an anisotropic crystal is detected through a linear polarizer while the crystal is rotated relative to the detection axis by angle $\varphi$.
\pnl{b} Schematic of the radially polarized transition-radiation field, where the detected field projects differently onto the two principal in-plane optical axes. 
\pnl{c-e} Angle-resolved isotropic transition radiation from vacuum-platinum interface, measured without a polarizer \pnl{c} and with the analyser at $\beta=0^\circ$ \pnl{d} and $\beta=90^\circ$ \pnl{e} (30\,keV, 1.4\,nA). Black regions indicate angles not collected by the mirror.
}
\label{fig-intro}
\end{figure}

\subsection*{Detection of anisotropic CL}

To experimentally access the anisotropic optical response of a biaxial crystal, we use polarization-selective detection to resolve the spectral contributions associated with its principal dielectric axes.
Fig.~\ref{fig-intro}\pnl{a} shows the polarization-resolved CL geometry used here.
A focused electron beam excites transition radiation in the suspended crystal, while the emitted light is collected by an aluminium parabolic mirror, passed through a linear polarizer acting as an analyser, and characterized with a spectrometer (see Methods).

Fig.~\ref{fig-intro}\pnl{b} illustrates the radial polarization of transition radiation, for which different emission directions contain different field components along the two principal in-plane optical axes.
In the parabolic-mirror geometry commonly used in CL spectroscopy, however, the detected polarization is also affected by the mirror itself.
Reflection from the curved mirror modifies the polarization in an angle-dependent manner because different emission directions encounter different local $s$- and $p$-polarized reflection conditions~\cite{Coenen2012}.
As a result, a linear analyser does not simply select a polarization component, but also weights different parts of the angular emission pattern depending on its orientation.
Consequently, rotating the analyser changes the angular distribution contributing to the detected spectrum and can introduce spectral variations that are unrelated to the intrinsic anisotropy of the crystal.

We illustrate this effect experimentally in Fig.~\ref{fig-intro}\pnl{c-e} using transition radiation from an isotropic platinum surface.
Without a polarizer, the characteristic donut-shaped angular distribution of transition radiation from a dipole oriented normal to the interface is observed [Fig.~\ref{fig-intro}\pnl{c}].
With the analyser at $\beta=0^\circ$, emission along the mirror centreline is transmitted [Fig.~\ref{fig-intro}\pnl{d}], whereas it is strongly suppressed for $\beta=90^\circ$ [Fig.~\ref{fig-intro}\pnl{e}].
This behaviour follows from the radial polarization of transition radiation and the angle-dependent polarization transformation introduced by the parabolic mirror.
Rotating the analyser would therefore vary both the detected polarization and the angular weighting of the collected emission, making a direct comparison between the two principal optical axes ambiguous.
We instead keep the analyser fixed at $\beta=0^\circ$ and rotate the sample by an angle $\varphi$, thereby varying the orientation of the principal in-plane dielectric axes relative to a fixed detection polarization and collection geometry.
This approach extends parabolic-mirror CL to direct measurements of in-plane optical anisotropy.  

\begin{figure}
    \centering   \includegraphics[width=0.99\linewidth]{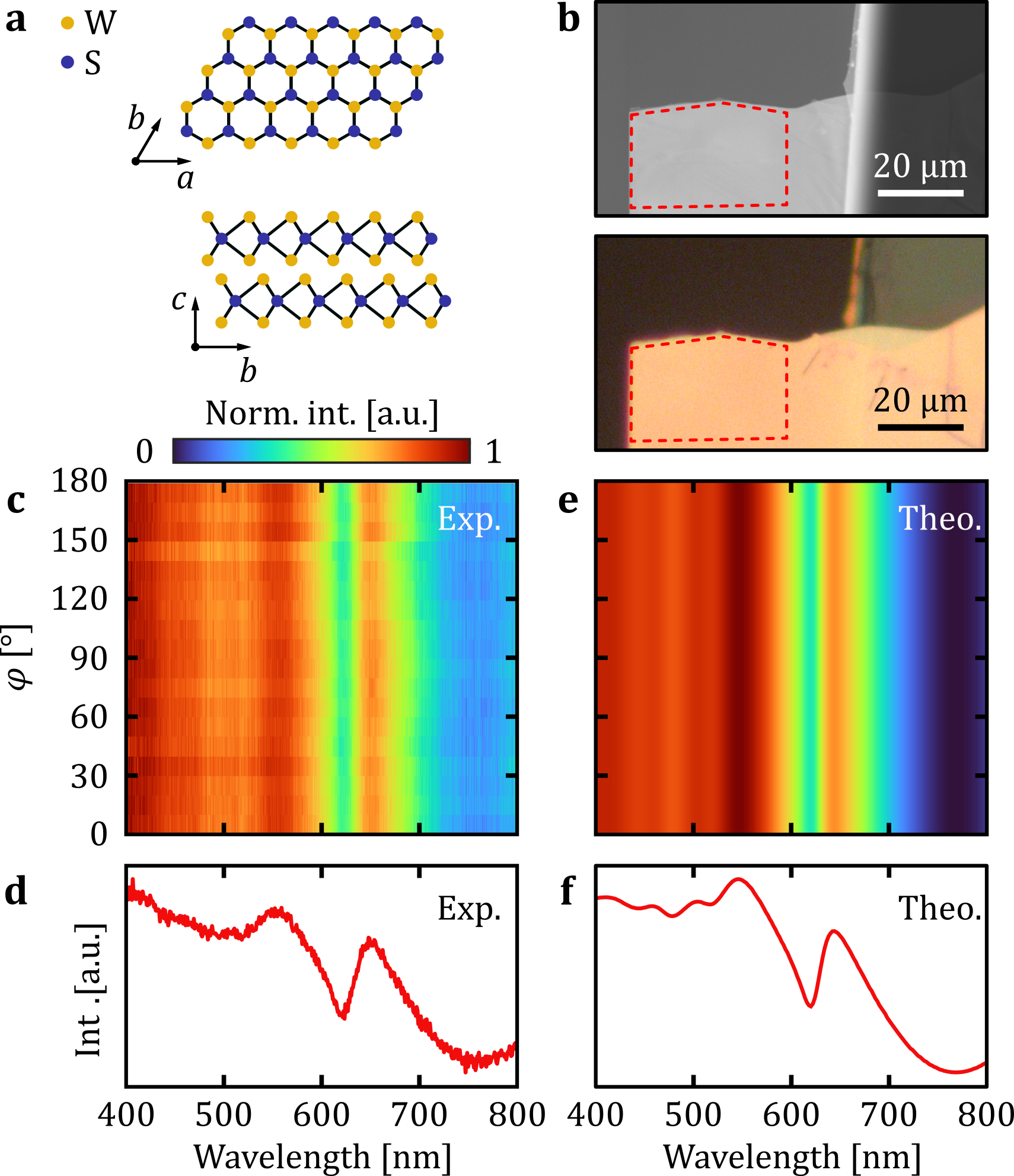}
    \caption{\textbf{Isotropic transition radiation from suspended WS$_2$ in vacuum.} 
   \pnl{a}~Crystal structure of 2H-WS$_2$.
\pnl{b} Scanning electron microscopy and optical images of the investigated 100\,nm-thick WS$_2$ flake; the red dashed outline marks the suspended region used for measurements, while the supporting grid can be seen on the right.
\pnl{c,e} Experimental and calculated polarization-resolved transition-radiation spectra as a function of in-plane sample rotation angle $\varphi$. 
\pnl{d,f} Corresponding spectral profiles acquired at \red{$\varphi=90^\circ$}. 
Experimental spectra were acquired at 30\,keV and 1.4\,nA with an integration time of 90\,s.}
    \label{WS2}
\end{figure}

\subsection{Isotropic transition radiation}

We first introduce transition radiation from an in-plane isotropic crystal as a benchmark. In this case, the two principal in-plane dielectric tensor components are equivalent, so rotating the crystal about the surface normal does not change the dielectric response relative to the fixed analyser. The measured transition-radiation spectrum is therefore expected to remain independent of the in-plane rotation angle $\varphi$. 

As a model system, we use WS$_2$, a layered van~der~Waals semiconductor with the hexagonal crystal structure shown in Fig.~\ref{WS2}\pnl{a}. Owing to its in-plane symmetry, 2H-WS$_2$ can be treated as optically isotropic within the basal plane. We investigate a suspended 100\,nm-thick WS$_2$ flake, shown in the scanning electron microscopy and optical images in Fig.~\ref{WS2}\pnl{b}.
The investigated crystal is suspended to provide two well-defined, parallel vacuum--material interfaces, enabling transition radiation from the entrance and exit surfaces to interfere without additional substrate contributions~\cite{Ebel_TREX_2026}.

Fig.~\ref{WS2}\pnl{c} shows the experimental polarization-selective spectra acquired while rotating the sample in $\Delta\varphi=10^\circ$ steps, while a representative spectral profile at \red{$\varphi=90^\circ$} is shown in Fig.~\ref{WS2}\pnl{d}. 
Although CL can contain both coherent and incoherent contributions, the spectral features observed here are assigned to transition radiation, as discussed in detail in Supplementary Note~\red{S1}.
The main spectral features remain at fixed wavelengths and exhibit no systematic dependence on the sample orientation, consistent with the in-plane isotropic dielectric response of WS$_2$.
To compare the measurements with theory, we calculate the transition-radiation response using the thin-film model adopted from our previous work~\cite{Ebel_TREX_2026} \red{(Methods and SI)}. In this description, transition radiation generated when the electron crosses the two vacuum--WS$_2$ interfaces interferes within the finite-thickness film, producing wavelength-dependent spectral maxima and minima. The dielectric function of WS$_2$ is taken from Ref.~\cite{Munkhbat2022}, while the film thickness of 100~nm was measured by optical reflection spectroscopy (Methods). The calculated spectra are shown in Fig.~\ref{WS2}\pnl{e} and, as expected for an in-plane isotropic dielectric response, are independent of $\varphi$.
Representative experimental and calculated spectra at \red{$\varphi=90^\circ$} are shown in Fig.~\ref{WS2}\pnl{d} and \pnl{f}, respectively. 
The simulations reproduce the characteristic spectral shape observed experimentally, including the pronounced features arising from the interaction of the thin-film transition-radiation resonances with the excitonic dielectric response of WS$_2$. 
We confirm our results using another isotropic vdW material, GaS, for which the measured $\varphi$-independent transition-radiation spectra are also in excellent agreement with the corresponding calculations (see Supplementary Note~\red{S2}).
Therefore, using the polarization-selective CL geometry introduced in Fig.~\ref{fig-intro}, we confirm the expected in-plane isotropic transition-radiation response of WS$_2$, with no systematic spectral modulation upon sample rotation.

\begin{figure}
    \centering       \includegraphics[width=0.99\linewidth]{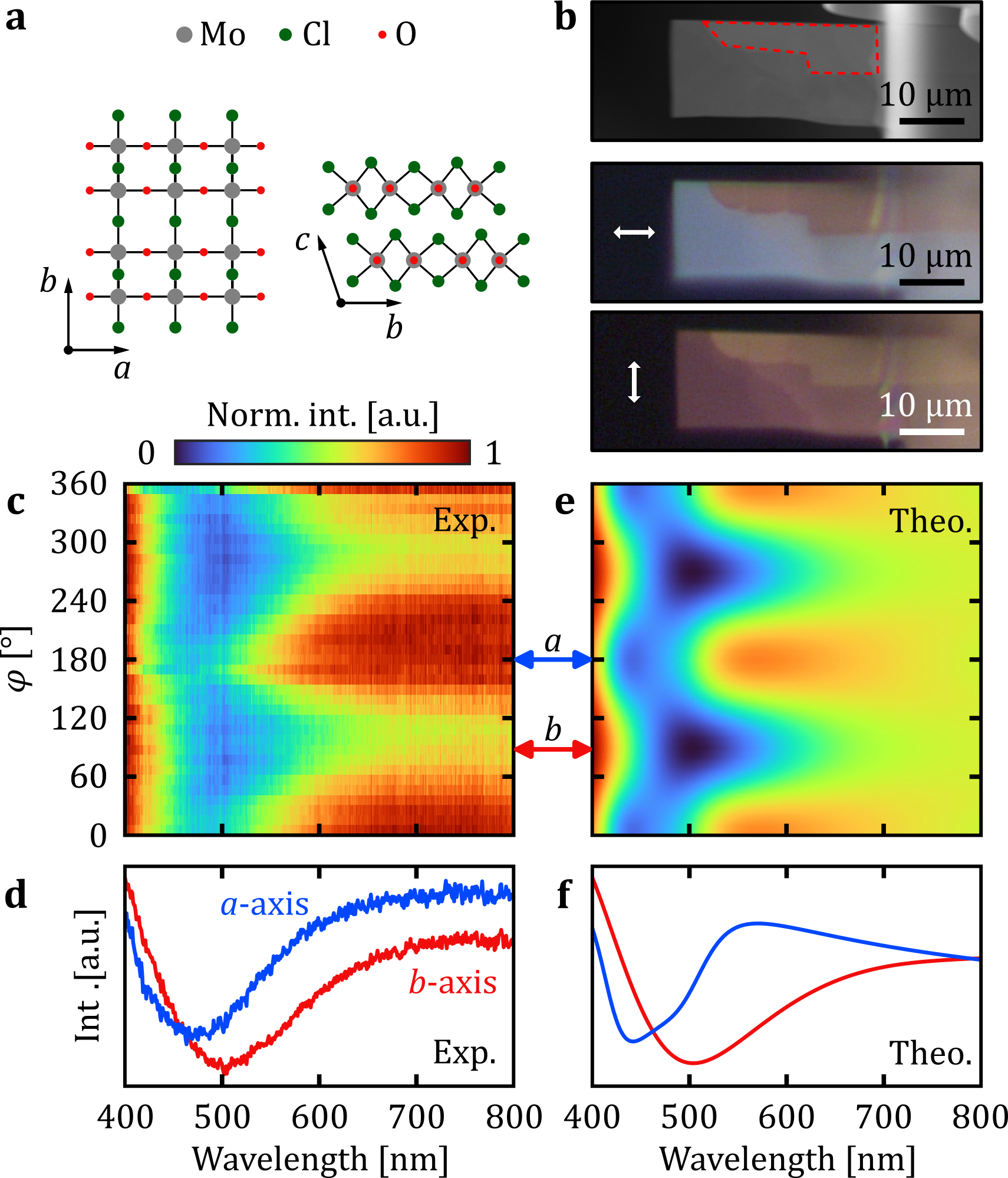}
\caption{\textbf{Anisotropic transition radiation from suspended MoOCl$_2$ in vacuum.} 
\pnl{a}~Crystal structure of MoOCl$_2$, showing the the crystallographic axes. 
\pnl{b}~Secondary electron and optical image of the investigated crystal. The red dashed outline indicates the suspended region used for the measurements, while the supporting grid can be seen on the right. 
The white arrows in optical images indicate the orientation of the incident light polarization. 
\pnl{c,e} Experimental and calculated polarization-resolved transition-radiation spectra as a function of in-plane sample rotation angle $\varphi$. 
\pnl{d,f} Corresponding spectral profiles acquired for the principal $a$- and $b$-axes at {$\varphi=180^\circ$} and {$\varphi=90^\circ$}, respectively. 
Experimental spectra were acquired at 30\,keV and 1.4\,nA with an integration time of 90\,s.
}
\label{MoOCl2}
\end{figure}

\subsection*{Anisotropic transition radiation}

With the isotropic response serving as a reference, we now investigate how transition radiation is modified by strong in-plane optical anisotropy. For this purpose, we selected MoOCl$_2$, a layered oxychloride with pronounced in-plane optical anisotropy. As illustrated in Fig.~\ref{MoOCl2}\pnl{a}, the crystal structure contains strongly coupled Mo--O chains along one in-plane direction, while the coupling along the orthogonal direction is significantly weaker. This structural anisotropy gives rise to very different optical responses along the crystallographic $a$- and $b$-axes, with MoOCl$_2$ exhibiting metallic and dielectric behaviour along the two orthogonal in-plane directions over the visible and near-infrared spectral range~\cite{Ruta2025}.
A suspended MoOCl$_2$ flake with a thickness of 85\,nm is shown in Fig.~\ref{MoOCl2}\pnl{b}. 
The top panel shows a secondary-electron image with the measured region indicated. Polarized reflection images reveal a pronounced colour contrast between orthogonal polarizations, manifesting the strong in-plane optical anisotropy and allowing the crystallographic axes to be identified prior to the CL measurements~\cite{Li2025}.

We next repeat the polarization-selective CL measurements while rotating the MoOCl$_2$ crystal relative to the fixed analyser [Fig.~\ref{MoOCl2}\pnl{c}]. In contrast to the isotropic case, the spectrum changes periodically with crystal orientation, directly demonstrating that the transition-radiation response depends on the in-plane optical direction. The strongest modulation occurs around the pronounced spectral minima near 470--500\,nm, whose positions and amplitudes vary as the detected polarization is rotated relative to the crystallographic axes. The corresponding spectra along the principal $a$- and $b$-axes, extracted at $\varphi=180^\circ$ and $\varphi=90^\circ$, respectively, are shown in Fig.~\ref{MoOCl2}\pnl{d} and exhibiting distinct spectral responses. 

To describe this behaviour, we introduce the anisotropic transition-radiation model to include the two principal in-plane dielectric directions (Methods). For a freely suspended film of thickness $d$, transition radiation generated at the two vacuum--material interfaces interferes, and the resulting intensity is calculated as $I(\theta,\lambda,d,\beta)$, where $\theta$ is the emission angle with respect to the normal to sample plane, and $\beta=v_{\mathrm{el}}/c$ \cite{Ebel_TREX_2026}. For the biaxial case, we evaluate this response separately using the two in-plane dielectric functions.
Because transition radiation generated under normal electron incidence is radially polarized, the detected linear polarization projects differently onto the two in-plane optical axes. Within this approximation, the detected anisotropic response is written as
\begin{equation}\label{TR_aniso_projection}
\begin{aligned}
I_{\mathrm{aniso}}(\theta,\lambda,d,\beta,\varphi)
=
&I(\theta,\lambda,d,\beta;\varepsilon_{xx})\cos^2\varphi \\
&+
I(\theta,\lambda,d,\beta;\varepsilon_{yy})\sin^2\varphi .
\end{aligned}
\end{equation}
This principal-axis projection model does not represent a full tensorial treatment of transition radiation in a biaxial slab, but provides an effective description of the measured spectra in terms of the two inequivalent in-plane dielectric responses.

The calculated angular--spectral map is shown in Fig.~\ref{MoOCl2}\pnl{e}. It reproduces the periodic dependence on $\varphi$ and the main spectral features observed experimentally. The corresponding calculated principal-axis spectra in Fig.~\ref{MoOCl2}\pnl{f} reproduce the distinct minima associated with the $a$- and $b$-axis responses, showing that the model captures the dominant effect of the anisotropic dielectric response on transition radiation.
We confirm this behaviour in another anisotropic vdW material, GeS, for which the measured $\varphi$-dependent transition-radiation spectra are also in excellent agreement with the corresponding calculations (see Supplementary Note~\red{S3}). 
Together, these measurements show that transition radiation can resolve the inequivalent in-plane optical responses of anisotropic vdW materials.

\subsection*{Electron-energy control}

Finally, we demonstrate that the transition-radiation resonances associated with the two principal crystallographic axes can be independently tuned by varying the electron-beam energy. Because transition radiation is generated at both the entrance and exit interfaces, the resulting spectrum is governed by the relative phase between the two contributions, which depends on the electron velocity as well as on the optical phase accumulated within the film~\cite{Ebel_TREX_2026}. Changing the electron energy therefore shifts the transition-radiation resonances.

\begin{figure}
    \centering   \includegraphics[width=0.99\linewidth]{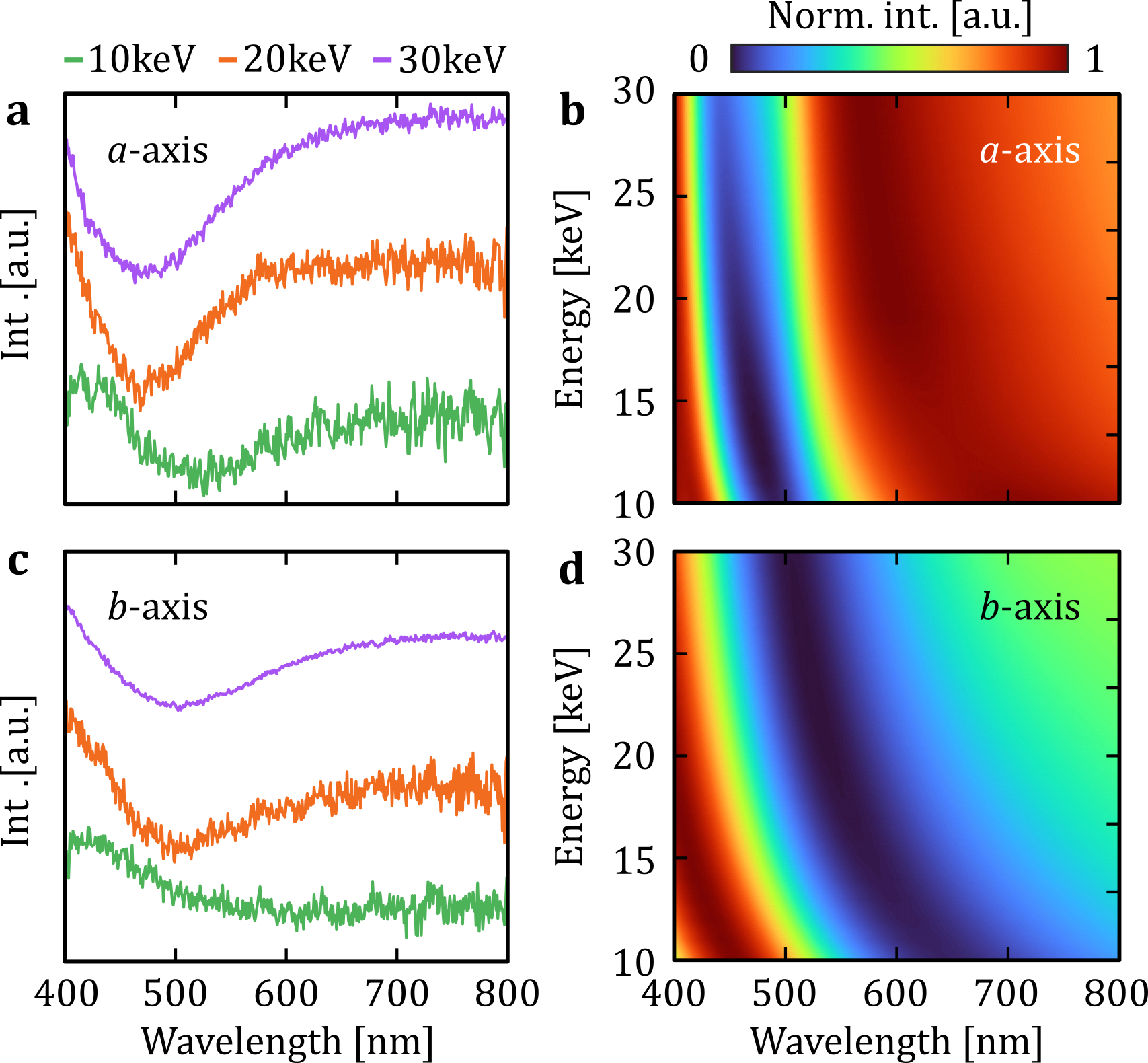}
   \caption{\textbf{Electron-energy tuning of anisotropic transition-radiation in MoOCl$_2$.}
Experimental and calculated transition-radiation spectra for the crystallographic as a function of electron-beam energy for \pnl{a-b} $a$-axis and \pnl{b-c} $b$-axis. Experimental spectra were acquired at a beam current of 1.4\,nA with an integration time of 90\,s.
}
    \label{Velocity}
\end{figure}

To investigate anisotropy of this effect, we measured the 85\,nm thick MoOCl$_2$ crystal at electron energies of 10, 20, and 30\,keV. Fig.~\ref{Velocity}\pnl{a} shows the experimental spectra measured along the crystallographic $a$-axis. The spectral minimum shifts with decreasing electron energy, while the overall response remains comparatively weakly dispersive. The calculated response in Fig.~\ref{Velocity}\pnl{b} reproduces this behaviour and shows the evolution of the transition-radiation spectrum over the full 10--30\,keV range.

A distinctly different dependence is observed along the crystallographic $b$-axis. The experimental spectra in Fig.~\ref{Velocity}\pnl{c} show a stronger shift of the spectral minimum as the electron energy is reduced. At 10\,keV, the minimum moves toward longer wavelengths and approaches the edge of the measured spectral range. The corresponding calculations in Fig.~\ref{Velocity}\pnl{d} reproduce this stronger electron-energy dependence.
The different tuning behaviour along the two crystallographic axes originates from  the  different in-plane dielectric responses of MoOCl$_2$. 
This makes the spectral response of transition radiation not only orientation dependent, but also independently tunable along each crystallographic direction through the electron energy, providing an additional handle to control anisotropic transition radiation.

\section*{Conclusion and outlook}

In conclusion, we have introduced a polarization-selective CL for probing in-plane optical anisotropy and applied it to transition radiation from suspended vdW crystals. Measurements on the in-plane isotropic crystals WS$_2$ and GaS confirm the expected orientation-independent response, whereas MoOCl$_2$ and GeS exhibit pronounced orientation-dependent spectra arising from their inequivalent in-plane dielectric responses. These measurements reveal spectrally anisotropic transition radiation and show that its resonances can be selectively tuned along the different crystallographic axes by varying the electron energy. The observed behaviour is reproduced by an effective thin-film model that relates the observed spectra to the principal in-plane dielectric functions. The present model treats the two in-plane dielectric axes independently and combines their transition-radiation responses through polarization projection. While this captures the dominant spectral anisotropy, a complete description will require a fully vectorial treatment including the full dielectric tensor, field-component coupling, and polarization conversion at anisotropic interfaces.

The sensitivity of transition radiation to \red{the tensor form of} the dielectric response suggests a route to transition-radiation spectroscopy: using the spectral, angular, and polarization properties of electron-induced transition radiation to probe the local optical response of a material. This could enable nanoscale determination of principal optical-axis orientations and spatially resolved mapping of variations in the dielectric function arising from strain, crystal domains, composition, carrier density, or structural phase changes~\cite{Rich1997,Ermolaev2024,Akbari2022}. 
Beyond spectroscopy, material anisotropy provides additional polarization and orientation control over transition radiation, with applications in free-electron light sources, particle detection, and electron-beam diagnostics~\cite{Happek1991,vanTilborg2006}.
Finally, the polarization-selective CL geometry is not restricted to transition radiation. The electron beam can excite out-of-plane dipoles, while polarization-selective detection can isolate their contribution, enabling studies of optical excitations that couple only weakly to conventional far-field illumination. 
\section*{Methods}
\noindent \textbf{Transition radiation model.} 
The transition-radiation intensity $I(\theta,\lambda,d,\beta)$ from an isotropic thin film can be described using the analytical model developed by Ritchie and Eldridge~\cite{Ritchie1962}.  For normal electron-beam incidence, the intensity per incident electron emitted into a unit solid angle at polar angle $\theta$, per unit wavelength $\lambda$, and in the backward direction with respect to the electron trajectory is given by
\begin{subequations}
\begin{equation}\label{TR_model}
I(\theta,\lambda,d,\beta)=\frac{\alpha\beta^2}{\pi^2\lambda}\mu^2(1-\mu^2)\left|\frac{\gamma}{\Delta}\right|^2 ,
\end{equation}
\begin{equation}\label{Term_contr.}
\begin{aligned}
\gamma = 
&\left[\frac{\varepsilon - \beta\sigma}{1 - \mu^2\beta^2}- \frac{1}{1 + \beta\sigma}
\right](\mu\varepsilon + \sigma) \exp[-i t \sigma]
\\[4pt]
&+
\left[-\frac{\varepsilon + \beta\sigma}{1 - \mu^2\beta^2}+ \frac{1}{1 - \beta\sigma}\right](\mu\varepsilon - \sigma) \exp[i t \sigma]
\\[4pt]
&-2\sigma\left[\frac{\varepsilon}{1+\beta\mu}-\frac{1-\beta\varepsilon\mu}{1-\beta^2\sigma^2}\right] \exp[i t/\beta]\,,
\end{aligned}
\end{equation}
\begin{equation}
\Delta=(\mu\varepsilon-\sigma)^2\exp[it\sigma]-(\mu\varepsilon+\sigma)^2 \exp[-it\sigma]\,,
\end{equation}
\end{subequations}
where $c$ is the speed of light, $\alpha$ is the fine-structure constant, $\sigma=(\varepsilon-1+\mu^2)^{1/2}$, $\beta=v_{\mathrm{el}}/c$, $\mu=\cos\theta$, and $t=2\pi d/\lambda$.
For in-plane anisotropic materials, the same expression is evaluated separately for the two principal in-plane dielectric functions, $\varepsilon_{xx}$ and $\varepsilon_{yy}$, yielding the corresponding principal-axis transition-radiation spectra. Polarization-resolved spectra are then calculated by inserting these principal-axis intensities into the projection model defined in Eq.~\eqref{TR_aniso_projection}. For in-plane isotropic materials, the two in-plane dielectric functions are identical, $\varepsilon_{xx}=\varepsilon_{yy}$, such that Eq.~\eqref{TR_aniso_projection} becomes independent of the in-plane rotation angle.

\noindent \textbf{CL measurements.}
CL measurements were performed using a TESCAN MIRA3 scanning-electron microscope (SEM) equipped with a Schottky field-emission gun and a SPARC Spectral detection system (Delmic). The setup includes an aluminium-coated parabolic mirror positioned above the sample stage, with a central aperture that allows the electron beam to pass through while collecting the generated CL. The mirror provides an effective numerical aperture of 0.97, and the collected light is directed through a Glan--Thompson polarizer (Thorlabs, GTH5) to a spectrograph (Andor Kymera 193i) equipped with a 150~lines/mm grating blazed at 500\,nm and detected using a CCD camera (Andor Newton). Measurements were performed at acceleration voltages between 10 and 30\,kV using a beam current of 1.4\,nA, measured with a Faraday cup. Spectra were acquired with exposure times of 90--120\,s and a beam spot size of approximately 10\,nm to ensure sufficient signal-to-noise ratio. Angle-resolved CL maps were recorded with exposure times of 90\,s. All CL spectra and angular maps were background-corrected using dark counts acquired with the electron beam blanked and subsequently corrected for the spectral response function of the instrument which was determined including the Glan--Thompson polariser~\cite{Ebel2025}.

\noindent \textbf{Sample fabrication.}
The WS$_2$, GeS, and GaS crystals were purchased from HQ Graphene. 
Samples were prepared by mechanical exfoliation onto polydimethylsiloxane (PDMS) stamps and subsequently transferred by deterministic viscoelastic stamping~\cite{CastellanosGomez2014}. Silicon wafers covered with silicon nitride (Si$_3$N$_4$) forming TEM windows were used as target substrates. Prior to transfer, the Si$_3$N$_4$ membrane was removed from the window area, allowing the selected van~der~Waals crystals to be freely suspended while remaining supported at the window edges. During transfer, the substrate was heated to 110$^\circ$C to promote release of the flake from the PDMS stamp. The thicknesses of the suspended crystals were determined by optical reflection spectroscopy, as described in Supplementary Note~S2.

\section*{Author contributions}

S.~E. conceived the idea. M.~S.~J provided MoOCl$_2$ crystals. The CL spectroscopy was done by S.~E. and S.~M., while spectroscopic imaging ellipsometry was performed by B.~M.~F., S.~E., and M.~B.-G. The work was supervised by S.~M., M.~B.-G., and N.~A.~M. All authors participated in data analysis and writing of the manuscript.

\section*{Acknowledgments}

The Center for Polariton-driven Light--Matter Interactions (POLIMA) is funded by the Danish National Research Foundation (Project No.~DNRF165).
S.~E. acknowledges funding from Thomas B. Thriges Fond for his research stay at the Heriot--Watt University. M~.B.-G. is supported by a Royal Society University Research Fellowship. B.~D.~G. is supported by a Chair in Emerging Technology from the Royal Academy of Engineering.

\bibliography{bibliography}

@article{Akbari2022,
  title = {Directional effects in plasmon excitation and transition radiation from an anisotropic {2D} material induced by a fast charged particle},
  volume = {14},
  ISSN = {2040-3372},
  url = {http://dx.doi.org/10.1039/d1nr06307c},
  DOI = {10.1039/d1nr06307c},
  number = {13},
  journal = {Nanoscale},
  publisher = {Royal Society of Chemistry (RSC)},
  author = {Akbari,  K. and Mišković,  Z. L.},
  year = {2022},
  pages = {5079–5093}
}

@article{Ermolaev2024,
  title = {Wandering principal optical axes in {van der Waals} triclinic materials},
  volume = {15},
  DOI = {10.1038/s41467-024-45266-3},
  pages = {1552},
  journal = {Nature Communications},
  publisher = {Springer Science and Business Media LLC},
  author = {Ermolaev,  G. A. and Voronin,  K. V. and Toksumakov,  A. N. and Grudinin,  D. V. and Fradkin,  I. M. and Mazitov,  A. and Slavich,  A. S. and Tatmyshevskiy,  M. K. and Yakubovsky,  D. I. and Solovey,  V. R. and Kirtaev,  R. V. and Novikov,  S. M. and Zhukova,  E. S. and Kruglov,  I. and Vyshnevyy,  A. A. and Baranov,  D. G. and Ghazaryan,  D. A. and Arsenin,  A. V. and Martin-Moreno,  L. and Volkov,  V. S. and Novoselov,  K. S.},
  year = {2024}
}

@article{Rich1997,
  title = {Linearly polarized and time-resolved cathodoluminescence study of strain-induced laterally ordered {(InP)$_2$/(GaP)$_2$} quantum wires},
  volume = {81},
  url = {http://dx.doi.org/10.1063/1.365243},
  DOI = {10.1063/1.365243},
  number = {10},
  journal = {Journal of Applied Physics},
  publisher = {AIP Publishing},
  author = {Rich,  D. H. and Tang,  Y. and Lin,  H. T.},
  year = {1997},
  pages = {6837–6852}
}

@article{Abdi2025,
  title = {{2D} Borophene: In‐Plane Hyperbolic Polaritons in the Visible Spectral Range},
  volume = {35},
  DOI = {10.1002/adfm.202513016},
  number = {39},
  pages = {e13016},
  journal = {Advanced Functional Materials},
  author = {Abdi,  Y. and Taleb,  M. and Black,  M. and Hajibaba,  S. and Moayedi,  M. and Talebi,  N.},
  year = {2025}
}

@article{Ebel2025,
  title = {An atlas of photonic and plasmonic materials for cathodoluminescence microscopy},
  ISSN = {2192-8614},
  url = {http://dx.doi.org/10.1515/nanoph-2025-0135},
  DOI = {10.1515/nanoph-2025-0135},
  journal = {Nanophotonics},
  publisher = {Walter de Gruyter GmbH},
  author = {Ebel,  S. and Lebsir,  Y. and Yezekyan,  T. and Mortensen,  N. A. and Morozov,  S.},
  year = {2025},
  volume = {14},
  number = {15},
pages = {2647-2671}
}

@article{Polman2019,
  title = {Electron-beam spectroscopy for nanophotonics},
  volume = {18},
  ISSN = {1476-4660},
  url = {http://dx.doi.org/10.1038/s41563-019-0409-1},
  DOI = {10.1038/s41563-019-0409-1},
  number = {11},
  journal = {Nature Materials},
  publisher = {Springer Science and Business Media LLC},
  author = {Polman,  A. and Kociak, M. and García de Abajo,  F. J.},
  year = {2019},
  pages = {1158–1171}
}

@article{Ritchie1962,
  title = {Optical Emission from Irradiated Foils. {I}},
  author = {Ritchie, R. H. and Eldridge, H. B.},
  journal = {Physical Review},
  volume = {126},
  number = {6},
  pages = {1935--1947},
  numpages = {0},
  year = {1962},
  publisher = {American Physical Society},
  doi = {10.1103/PhysRev.126.1935},
  url = {https://link.aps.org/doi/10.1103/PhysRev.126.1935}
}

@article{Yamamoto1996,
    author = {Yamamoto, N. and Sugiyama, H. and Toda, A.},
    title = {Cherenkov and transition radiation from thin plate crystals detected in the transmission electron microscope},
    journal = {Proceedings of the Royal Society A},
    volume = {452},
    number = {1953},
    pages = {2279-2301},
    year = {1996},
    issn = {1364-5021},
    doi = {10.1098/rspa.1996.0122},
    url = {https://doi.org/10.1098/rspa.1996.0122},
    eprint = {https://royalsocietypublishing.org/rspa/article-pdf/452/1953/2279/998684/rspa.1996.0122.pdf},
}

@article{Munkhbat2022,
  title = {Optical Constants of Several Multilayer Transition Metal Dichalcogenides Measured by Spectroscopic Ellipsometry in the 300–1700 nm Range: High Index,  Anisotropy,  and Hyperbolicity},
  volume = {9},
  ISSN = {2330-4022},
  url = {http://dx.doi.org/10.1021/acsphotonics.2c00433},
  DOI = {10.1021/acsphotonics.2c00433},
  number = {7},
  journal = {ACS Photonics},
  publisher = {American Chemical Society (ACS)},
  author = {Munkhbat,  B. and Wróbel,  P. and Antosiewicz,  T. J. and Shegai,  T. O.},
  year = {2022},
  pages = {2398–2407}
}

@article{CastellanosGomez2014,
  title = {Deterministic transfer of two-dimensional materials by all-dry viscoelastic stamping},
  volume = {1},
  ISSN = {2053-1583},
  url = {http://dx.doi.org/10.1088/2053-1583/1/1/011002},
  DOI = {10.1088/2053-1583/1/1/011002},
  number = {1},
  journal = {2D Materials},
  publisher = {IOP Publishing},
  author = {Castellanos-Gomez,  A. and Buscema,  M. and Molenaar,  R. and Singh,  V. and Janssen,  L. and van der Zant,  H. S. J. and Steele,  G. A.},
  year = {2014},
  pages = {011002}
}

@article{Brenny2014,
  title = {Quantifying coherent and incoherent cathodoluminescence in semiconductors and metals},
  volume = {115},
pages = {244307},
  ISSN = {1089-7550},
  url = {http://dx.doi.org/10.1063/1.4885426},
  DOI = {10.1063/1.4885426},
  number = {24},
  journal = {Journal of Applied Physics},
  publisher = {AIP Publishing},
  author = {Brenny,  B. J. M. and Coenen,  T. and Polman,  A.},
  year = {2014}
}

@article{Li2025,
  title = {Broadband near-infrared hyperbolic polaritons in {MoOCl$_2$}},
  volume = {16},
  url = {http://dx.doi.org/10.1038/s41467-025-61548-w},
  DOI = {10.1038/s41467-025-61548-w},
  pages = {6172},
  journal = {Nature Communications},
  author = {Li,  Y. and Zhang,  Y. and Zhang,  W. and Li,  X. and Tang,  J. and Xiao,  J. and Zhang,  G. and Liao,  X. and Jiang,  P. and Liu,  Q. and Luo,  Y. and Cao,  Z. and Lyu,  Q. and Tong,  Y. and Yang,  R. and Yang,  H. and Sun,  Q. and Gao,  Y. and Wang,  P. and Chen,  Z. and Liu,  W. and Wang,  S. and Lyu,  G. and Hu,  X. and Aeschlimann,  M. and Gong,  Q.},
  year = {2025}
}

@article{Ginzburg1996,
  title = {Radiation by uniformly moving sources ({Vavilov--Cherenkov} effect,  transition radiation,  and other phenomena)},
  volume = {39},
  ISSN = {1468-4780},
  url = {http://dx.doi.org/10.1070/PU1996v039n10ABEH000171},
  DOI = {10.1070/pu1996v039n10abeh000171},
  number = {10},
  journal = {Physics-Uspekhi},
  publisher = {Uspekhi Fizicheskikh Nauk (UFN) Journal},
  author = {Ginzburg,  V. L.},
  year = {1996},
  pages = {973–982}
}

@article{Lin2018,
  doi = {10.1038/s41567-018-0138-4},
  year = {2018},
  volume = {14},
  number = {8},
  author = {Lin,  X. and Easo,  S. and Shen,  Y. and Chen,  H. and Zhang,  B. and Joannopoulos,  J. D. and Soljačić,  M. and Kaminer,  I.},
  title = {Controlling {Cherenkov} angles with resonance transition radiation},
  pages = {816–821},
  journal = {Nature Physics},
}

@article{Coenen2012,
  title = {Polarization-sensitive cathodoluminescence {Fourier} microscopy},
  volume = {20},
  ISSN = {1094-4087},
  url = {http://dx.doi.org/10.1364/OE.20.018679},
  DOI = {10.1364/oe.20.018679},
  number = {17},
  journal = {Optics Express},
  publisher = {Optica Publishing Group},
  author = {Coenen,  T. and Polman,  A.},
  year = {2012},
  pages = {18679}
}

@book{BornWolf:1999:Book,
  asin = {0521642221},
  author = {Born, M. and Wolf, E.},
  description = {Amazon.com: Principles of Optics: Electromagnetic Theory of Propagation, Interference and Diffraction of Light: Max Born, Emil Wolf: Books},
  dewey = {535},
  ean = {9780521642224},
  edition = {7th},
  isbn = {0521642221},
  publisher = {Cambridge University Press},
  address = {Cambridge, UK},
  title = {Principles of Optics: Electromagnetic Theory of Propagation, Interference and Diffraction of Light},
  year = 1999
}

@book{LandauLifshitz1984,
  title     = {Electrodynamics of Continuous Media},
  author    = {Landau, L. D. and Lifshitz, E. M. and Pitaevskii, L. P.},
  edition   = {2nd},
  publisher = {Pergamon Press},
  year      = {1984}
}

@article{Chen2023,
  title = {Recent advances of transition radiation: Fundamentals and applications},
  volume = {3},
  ISSN = {2772-9494},
  url = {http://dx.doi.org/10.1016/j.mtelec.2023.100025},
  DOI = {10.1016/j.mtelec.2023.100025},
  journal = {Materials Today Electronics},
  publisher = {Elsevier BV},
  author = {Chen,  R. and Gong,  Z. and Chen,  J. and Zhang,  X. and Zhu,  X. and Chen,  H. and Lin,  X.},
  year = {2023},
  pages = {100025}
}

@article{Happek1991,
  title = {Observation of coherent transition radiation},
  volume = {67},
  ISSN = {0031-9007},
  url = {http://dx.doi.org/10.1103/PhysRevLett.67.2962},
  DOI = {10.1103/physrevlett.67.2962},
  number = {21},
  journal = {Physical Review Letters},
  publisher = {American Physical Society (APS)},
  author = {Happek,  U. and Sievers,  A. J. and Blum,  E. B.},
  year = {1991},
  pages = {2962–2965}
}

@article{vanTilborg2006,
  title = {Temporal Characterization of Femtosecond Laser-Plasma-Accelerated Electron Bunches Using Terahertz Radiation},
  volume = {96},
  pages = {014801},
  DOI = {10.1103/physrevlett.96.014801},
  number = {1},
  journal = {Physical Review Letters},
  author = {van Tilborg,  J. and Schroeder,  C. B. and Filip,  C. V. and Tóth,  Cs. and Geddes,  C. G. R. and Fubiani,  G. and Huber,  R. and Kaindl,  R. A. and Esarey,  E. and Leemans,  W. P.},
  year = {2006},
}

@article{RoquesCarmes2023,
  title   = {Free-electron--light interactions in nanophotonics},
  author  = {Roques-Carmes, C. and Kooi, S. E. and Yang, Y. and Rivera, N. and Keathley, P. D. and Joannopoulos, J. D. and Johnson, S. G. and Kaminer, I. and Berggren, K. K. and Solja{\v c}i{\'c}, M.},
  journal = {Applied Physics Reviews},
  volume  = {10},
  number = {1},
  pages   = {011303},
  year    = {2023},
  doi     = {10.1063/5.0118096}
}

@article{Chen_PRL_2023,
  title = {Low-Velocity-Favored Transition Radiation},
  volume = {131},
  ISSN = {1079-7114},
  url = {http://dx.doi.org/10.1103/PhysRevLett.131.113002},
  DOI = {10.1103/physrevlett.131.113002},
  number = {11},
  pages = {113002},
  journal = {Physical Review Letters},
  publisher = {American Physical Society (APS)},
  author = {Chen,  J. and Chen,  R. and Tay,  F. and Gong,  Z. and Hu,  H. and Yang,  Y. and Zhang,  X. and Wang,  C. and Kaminer,  I. and Chen,  H. and Zhang,  B. and Lin,  X.},
  year = {2023},
}

@article{Wang2025,
  title = {Ultra‐Directional Transition Radiation From Deep‐Subwavelength Epsilon‐Near‐Zero Metamaterials},
  volume = {14},
  ISSN = {2195-1071},
  url = {http://dx.doi.org/10.1002/adom.202501449},
  DOI = {10.1002/adom.202501449},
  number = {2},
  pages = {e01449},
  journal = {Advanced Optical Materials},
  publisher = {Wiley},
  author = {Wang,  Z. and Gong,  Z. and Chen,  R. and Xi,  X. and Chen,  J. and Yang,  Y. and Chen,  H. and Li,  E. and Kaminer,  I. and Lin,  X.},
  year = {2026},
}

@article{Chen_Brewster_2023,
  title = {Free-electron {Brewster}-transition radiation},
  volume = {9},
  ISSN = {2375-2548},
  url = {http://dx.doi.org/10.1126/sciadv.adh8098},
  DOI = {10.1126/sciadv.adh8098},
  number = {32},
  pages = {eadh8098},
  journal = {Science Advances},
  publisher = {American Association for the Advancement of Science (AAAS)},
  author = {Chen,  R. and Chen,  J. and Gong,  Z. and Zhang,  X. and Zhu,  X. and Yang,  Y. and Kaminer,  I. and Chen,  H. and Zhang,  B. and Lin,  X.},
  year = {2023},
}

@article{Yu2019,
  title = {Transition Radiation in Photonic Topological Crystals: Quasiresonant Excitation of Robust Edge States by a Moving Charge},
  volume = {123},
  ISSN = {1079-7114},
  url = {http://dx.doi.org/10.1103/PhysRevLett.123.057402},
  DOI = {10.1103/physrevlett.123.057402},
  number = {5},
  pages = {057402},
  journal = {Physical Review Letters},
  publisher = {American Physical Society (APS)},
  author = {Yu,  Y. and Lai,  K. and Shao,  J. and Power,  J. and Conde,  M. and Liu,  W. and Doran,  S. and Jing,  C. and Wisniewski,  E. and Shvets,  G.},
  year = {2019},
}

@article{Ebel_TREX_2026,
  doi = {10.1002/adfm.76737},
  author = {Ebel,  S. and Nørgaard,  M. and Nicolaisen Hansen,  C. and Mortensen,  N. A. and Morozov,  S.},
  title = {Velocity-tunable exciton-photon hybridization in cathodoluminescence},
  journal = {Advanced Functional Materials},
  volume = {36},
  number = {62},
  pages = {e76737},
  year = {2026}
}

@article{Ruta2025,
  title = {Good plasmons in a bad metal},
  volume = {387},
  ISSN = {1095-9203},
  url = {http://dx.doi.org/10.1126/science.adr5926},
  DOI = {10.1126/science.adr5926},
  number = {6735},
  journal = {Science},
  publisher = {American Association for the Advancement of Science (AAAS)},
  author = {Ruta,  F. L. and Shao,  Y. and Acharya,  S. and Mu,  A. and Jo,  N. H. and Ryu,  S. H. and Balatsky,  D. and Su,  Y. and Pashov,  D. and Kim,  B. S. Y. and Katsnelson,  M. I. and Analytis,  J. G. and Rotenberg,  E. and Millis,  A. J. and van Schilfgaarde,  M. and Basov,  D. N.},
  year = {2025},
  pages = {786–791}
}

@article{Ma2018,
  title = {In-plane anisotropic and ultra-low-loss polaritons in a natural {van der Waals} crystal},
  volume = {562},
  ISSN = {1476-4687},
  url = {http://dx.doi.org/10.1038/s41586-018-0618-9},
  DOI = {10.1038/s41586-018-0618-9},
  number = {7728},
  journal = {Nature},
  publisher = {Springer Science and Business Media LLC},
  author = {Ma,  W. and Alonso-González,  P. and Li,  S. and Nikitin,  A. Y. and Yuan,  J. and Martín-Sánchez,  J. and Taboada-Gutiérrez,  J. and Amenabar,  I. and Li,  P. and Vélez,  S. and Tollan,  C. and Dai,  Z. and Zhang,  Y. and Sriram,  S. and Kalantar-Zadeh,  K. and Lee,  S.-T. and Hillenbrand,  R. and Bao,  Q.},
  year = {2018},
  pages = {557–562}
}
\bibliographystyle{unsrt}

\end{document}